\documentclass{iau} 
\usepackage{graphicx}
\def\la{\mathrel{\mathchoice 
{\vcenter{\offinterlineskip\halign{\hfil$\displaystyle##$\hfil\cr<\cr\sim\cr}}}
{\vcenter{\offinterlineskip\halign{\hfil$\textstyle##$\hfil\cr<\cr\sim\cr}}}
{\vcenter{\offinterlineskip\halign{\hfil$\scriptstyle##$\hfil\cr<\cr\sim\cr}}}
{\vcenter{\offinterlineskip\halign{\hfil$\scriptscriptstyle##$\hfil\cr<\cr\sim\cr}}}}}

\def\ga{\mathrel{\mathchoice 
{\vcenter{\offinterlineskip\halign{\hfil$\displaystyle##$\hfil\cr>\cr\sim\cr}}}
{\vcenter{\offinterlineskip\halign{\hfil$\textstyle##$\hfil\cr>\cr\sim\cr}}}
{\vcenter{\offinterlineskip\halign{\hfil$\scriptstyle##$\hfil\cr>\cr\sim\cr}}}
{\vcenter{\offinterlineskip\halign{\hfil$\scriptscriptstyle##$\hfil\cr>\cr\sim\cr}}}}}
\title[Black hole astrophysics with HAWC] %% give here short title %%
{Black hole astrophysics with HAWC,\\ the High Altitude Water Cherenkov \\ $\gamma$-ray observatory} 
\author[Carrami\~nana]   %% give here short author list %%
{Alberto Carrami\~nana$^1$
 {\em for the HAWC Collaboration}}

\affiliation{$^1$Instituto Nacional de Astrof\'{\i}sica, \'Optica y Electr\'onica\\ Luis Enrique Erro 1, Tonantzintla, Puebla, M\'exico\\email: {\tt alberto@inaoep.mx}}

\pubyear{2016}
\volume{324}  %% insert here IAU Symposium No.
\jname{New Frontiers in Black Hole Astrophysics}
\editors{A.C. Editor, B.D. Editor \& C.E. Editor, eds.}
\begin{document}

\maketitle

\begin{abstract}
The HAWC gamma-ray observatory is a wide field of view and high duty cycle $\gamma$-ray detector investigating the 0.1 - 100 TeV energy range. It has detected supermassive black holes in the near Universe, and is seeking to detect black hole related objects like gamma-ray bursts, Galactic binary systems, primordial black holes and gravitational wave mergers. Daily light curves of the BL Lac objects Mrk~421 and Mrk~501 are presented here, together with a compilation of studies of black hole related objects.
\end{abstract}

\section{Introduction: investigating black holes with HAWC}
%----------------------------------------------------------------------------%
Black holes, an early prediction of general relativity, are objects of extreme astrophysical interest (\cite{blackholes1}; \cite{blackholes2}). They can exist in radically different mass regimes, having originated in very different processes: primordial black holes ranging from the mass of a mountain, formed in the first instants of the Universe; stellar mass black holes, found in binary systems, are products of the stellar evolution of massive stars; and supermassive black holes, believed to be the power engines of active galactic nuclei, are thought to have built up in the early stages of cosmological evolution. The existence of black holes was debated for decades, until the discovery of the extragalactic nature and rapid variability of active galactic nuclei~(\cite{bh-agn};~\cite{agn-variability}), the determination of the mass of the binary stellar system Cygnus X-1~(\cite{cyg-x1}), and the kinematic evidence of a $4\times 10^{6}~\rm M_\odot$ in the center of the Galaxy~(\cite{bh-gc}). Recently, the Laser Interferometer Gravitational-Wave Observatory (LIGO) provided unmistakable evidence of gravitational waves emitted from unexpected $\sim 10-30\,{\rm M}_\odot$ binary system mergers~(\cite{gw1-ligo}, 2016b).

For half a century, the strong gravity inherent to black holes prompted their potential as high energy sources~(\cite{bh-agn}). 
% early paper of BH as astrophysical sources
The recent availability of the High Altitude Water Cherenkov (HAWC) observatory, a wide-field of view and high duty cycle $\gamma$-ray TeV detector of unprecedented sensitivity, now provides unique opportunities for black holes astrophysics. HAWC has been monitoring two thirds of the sky every sidereal day since it started early operations in August 2013, with upgraded sensitivity to detect sources with the TeV flux of the Crab in a single transit once it reached full operations in March 2015. HAWC has studied blazars and very high energy $\gamma$-ray sources in the Galaxy for over a year. Its science case includes investigating TeV emission from gamma-ray bursts, primordial black holes and X-ray binary systems. In addition, HAWC is an ideal instrument for multi-messenger follow-ups of neutrino sources or for the search for electromagnetic counterparts of gravitational wave events.

\section{Gamma-ray detectors}
%--------------------------------------%
Gamma rays, the most extreme form of electromagnetic radiation, are produced in violent astrophysical phenomena. They have been observed 
% energies range from about one MeV 
up to at least 100 TeV, with searches conducted up to $10^{18}\,\rm eV$~(\cite{auger-limit}). The atmosphere of the Earth is opaque to $\gamma$ rays, prompting their observation with space borne observatories in the MeV and GeV ranges. Individual very high energy $\gamma$ rays, as well as cosmic rays with $E\ga 100~\rm GeV$, have a perceptible effect in the atmosphere, triggering secondary particle cascades that can be detected with ground based instruments. In consequence, the $\gamma$-ray region of the electromagnetic spectrum is nowadays investigated with three types of instruments of complementary capabilities.

\subsection{Space borne telescopes}
Gamma-ray space telescopes detect photons through either the Compton effect, at low MeV energies, or pair production, at higher energies, in the GeV band. These instruments have wide fields of view which allow for complete sky surveys.  Up to now, there has been only one sky survey in low energy $\gamma$ rays, specifically in the 0.7--30~MeV range by {\em CGRO}-COMPTEL, in operations between 1991 and 2000~(\cite{comptel-cat}). In contrast, several pair production telescopes have scanned the sky with increasing depth, with the Large Area Telescope (LAT) on-board the {\em Fermi $\gamma$-ray Space Telescope} detecting over 3000 sources, in the deepest sky survey from 100~MeV--300~GeV summarized in the 3FGL catalog~(\cite{3fgl}); furthemore, {\em Fermi}-LAT has extended its performance to the 50~GeV--2~TeV in the 2FHL catalog~(\cite{2fhl}). {\em Fermi}-LAT has confirmed blazars as the dominant type of celestial $\gamma$-ray source, as previously proved by {\em CGRO}-EGRET~(\cite{3eg}).

\subsection{Air Cherenkov telescopes} 
Air Cherenkov telescopes (ACTs) study the very high energy (VHE) range, from about 30~GeV to 30 TeV, using the atmosphere as part of the detector system. Most secondary charged particles produced in atmospheric cascades travel faster than the speed of light in the air, $v>c/n$, emitting Cherenkov radiation confined to a narrow $\sim 2^\circ$ cone. This radiation is detectable from the ground. Working on dark night-time background during cloudless nights, ACTs can perform deep observations of TeV point sources down to the flux of the Crab in less than a minute~(\cite{crab-hess}). ACTs have narrow fields of view and low duty cycle, as their operation is restricted to nights with clear and moonless conditions. Their high sensitivity, in particular for point sources, make them optimal for detailed studies of targeted objects, producing high resolution spectra and sub-arcminute source localizations. They can partially compensate their narrow field-of-view through their high sensitivity in order to perform dedicated surveys of areas of interest, covering $\la 1000~\rm deg^{2}$ over several years~(\cite{gps-hess}). Observations from different ACT observatories are summarized in the TeVCat, a heterogeneous but useful catalog of TeV $\gamma$-ray sources found with different TeV $\gamma$-ray observatories~(\cite{tevcat}).

\subsection{Extensive Air Shower arrays} 
Extensive air shower arrays (EAS) detect secondary particles reaching the detector array. They were developed for cosmic-ray studies in the 1950s, with water Cherenkov detectors in use already in the mid-1960s~(\cite{cr-atmospheric}; \cite{haverah-park}). Current EAS like HAWC and Tibet AS-$\gamma$, have the ability to separate $\gamma$-ray primaries from hadronic cosmic rays, previously developed by the MILAGRO Collaboration~(\cite{crab-milagro}). These instruments have large fields of view ($\la 2\,\rm sr$) and duty cycles close to 100\%, which permit surveys of $>50\%$ of the sky transiting through their zenith-pointed instantaneous fields of view. Ideally, they operate with very little interruption, performing very shallow daily unbiased surveys that build in depth as their data accumulate. In particular, HAWC reaches down to a $1\sigma$ sensitivity level of $\sim 0.2~{\rm Crab}\left(t/\rm day\right)^{-1/2}$. EAS are best suited for transient source monitoring, the study of extended diffuse emission regions and reaching high photon energies, up to 100~TeV. 

\section{The HAWC  $\gamma$-ray observatory and its data}
\subsection{The HAWC $\gamma$-ray observatory}
The HAWC observatory was installed and is currently operated by a collaboration of about thirty institutions of Mexico and United States, including over a hundred scientists, with the recent incorporation of the Max Planck Institute at Heidelberg, the Cracow University and the associate membership of the University of Costa Rica. 
HAWC is a second generation water Cherenkov observatory (WCO), built from the experience gained during the construction and operation of the MILAGRO observatory that operated in New Mexico, between 1999 and 2008. HAWC is located at a higher altitude (4100~m compared to 2600~m) and benefits of an improved design. It has a sampling area of $22 000\,\rm m^2$, four times larger than MILAGRO; a ten times larger muon detection area, compared to MILAGRO; and includes optical isolation of its detector elements. The result is a  $\gamma$-ray detector 15 times more sensitive than its predecessor. % MILAGRO. 

The HAWC site is located in the Northern slope of Volc\'an Sierra Negra, also named Tliltepetl, inside the Parque Nacional Pico de Orizaba, in the Mexican state of Puebla. The geographical coordinates of the site are latitude of $+18.99^\circ\rm N$ and longitude $97.31^\circ\rm W$. The more equatorial location of HAWC, relative to MILAGRO, allows for a coverage of two-thirds of the sky, reaching down to the declination of the Galactic Center. HAWC shares the basic infrastructure developed between 1998 and 2006 for the Gran Telescopio Milim\'etrico Alfonso Serrano (GTM; also known as LMT), located on the 4600~m top of Sierra Negra. The GTM/LMT access road, electrical power line and optical fiber were extended to the HAWC site in 2010 and the array platform was prepared in 2011. With project funding available, the installation of the water Cherenkov array started in 2012, finishing in December 2014. The complete HAWC $\gamma$-ray observatory was inaugurated in March 2015, and has been in full operation since.
% since, building from the Science operations started in August 2013, with one-third of the array functional.

HAWC consists of 300 individual water Cherenkov detectors (WCD). Each WCD is made of a cylinder of corrugated steel of 7.2~m diameter and 5~m height, with an opaque bladder inside holding 180,000 liters of water and isolating it from environmental light, during day and night. Four photo-multiplier tubes (PMT) are located on the interior base of every WCD: three $8"$ PMTs that were previously used in MILAGRO, and a central fourth $10"$ PMT added to the design through the support of LANL-DoE. The PMT system provides fast response to Cherenkov light with high quantum efficiency in the blue to ultraviolet. Every single PMT is connected to the central counting house, where the experiment is controlled, data are acquired and stored, and the calibration system is run. A network of 180~km of underground cables connects each WCD to the counting house.

\subsection{HAWC data} 
HAWC is continuously sampling the ensemble of almost 1200~PMT channels, seeking for multiple coincident signals to declare ``an event''. Data are reconstructed event per event, considering events with more than 30 channels hit, $\rm nHit > 30$. Relative timing allows to locate the arrival direction of primary cosmic and $\gamma$ rays in the sky. Shower fronts are fitted with sub-nanosecond residuals, leading to localizations of $\sim 0.25^\circ$ in the sky validated with observations of the Crab Nebula. A datum  includes the charge deposits on each PMT channel, used for energy estimations and for $\gamma$/hadron discrimination through algorithms based on core location and the topology of the deposits in the array - including muon identification.
HAWC registers over 20,000 cosmic rays per second, generating 2 TB of data per day - every day. The data are physically transported to the HAWC-MX datacenter, at the Instituto de Ciencias Nucleares of UNAM, then transferred through the Internet to the HAWC-US datacenter at the University of Maryland.

\begin{figure} \begin{center}
\includegraphics[width=\hsize]{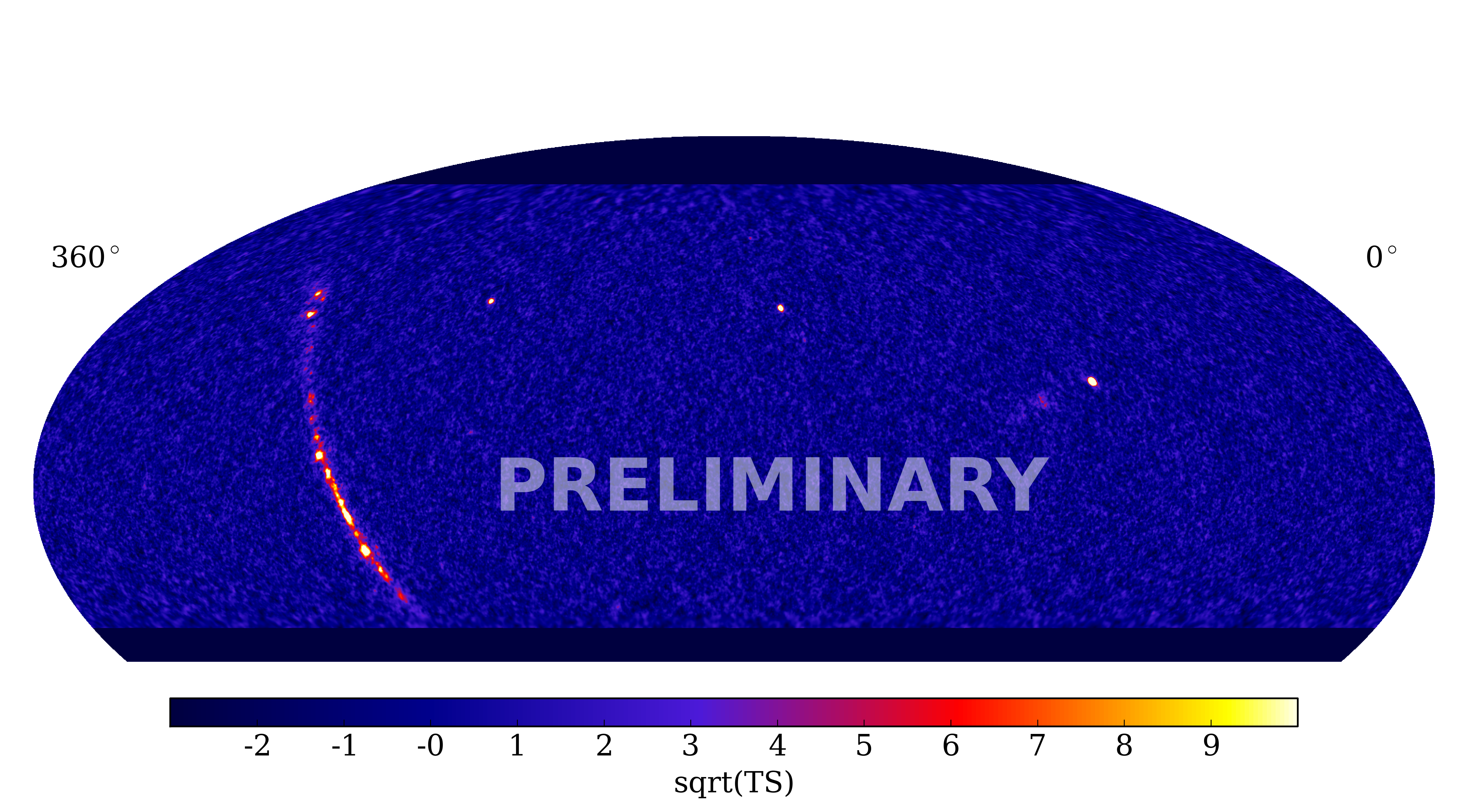}
\caption{Map of the sky seen in TeV $\gamma$ rays obtained with HAWC. The map is in celestial coordinates, covering  declinations between $+70^\circ$ and $-30^{\circ}$. The tilted emission band clearly visible on the left corresponds to the first Galactic Quadrant. The three outstanding point sources from left to right, Mrk~501, Mrk~421 and the Crab Nebula. Next to the Crab, with lower significance, is the extended emission coincident with Geminga.}\label{skymap}
\end{center} 
%\vspace*{-2mm}
\end{figure}

The HAWC Collaboration has analyzed the first 17~months of data. Sources have been searched for assuming point sources with spectral index $k=-2.7$. Three distinct features can be identified in the HAWC skymap, shown in Figure~\ref{skymap}: (1) the Galactic anti-center, highlighted by the Crab Nebula which is detected with a cumulative signal to noise of $\sim 100\sigma$ and is used to validate the performance of the observatory relative to Montecarlo simulations;
(2) two distinct active galaxies, Mrk~421 and Mrk~501, detected with sufficient strength as to perform daily monitoring of their TeV $\gamma$-ray fluxes; (3) the Galactic Plane, specifically its 1$^{st}$ Quadrant, {\em i.e.} longitudes $0 \la \ell\la 90^\circ$, where HAWC has detected over forty sources.

\section{Black hole astrophysics with HAWC}

\subsection{HAWC observations of supermassive black holes}
Supermassive black holes are currently believed to exist in the center of all galaxies. When accreting significant amounts of matter they power the nuclei of active galaxies, which in turn can produce cosmic and $\gamma$ rays in powerful, relativistic, jets. {\em CGRO}-EGRET and {\em Fermi}-LAT observations established BL Lacs and Flat Spectrum Radio Quasars (FSRQ) as the dominant types of GeV sources, several of them confirmed to extend their emission in the TeV range by ACTs. HAWC has made clear detections of Mrk~421 and Mrk~501, the two closest BL Lac objects known, in the 17-month data taken between November 2014 and February 2016. These are two of only four sources detected in the 0.6-2.0~TeV band by {\em Fermi}-LAT out of the Galactic plane, $|b|>5^\circ$, and at low redshift, $z<0.3$~(\cite{2fhl}). HAWC detected their emission with significances of 33$\sigma$ and 23$\sigma$, respectively, allowing for their daily monitoring.

Mrk~421 is the nearest BL Lac object known, at $d_{L}=134~Mpc$, as estimated from its redshift, $z=0.031$. Mrk~421 was originally detected at GeV energies in 1991-1992, during the early observations of {\em CGRO}-EGRET~(\cite{grocat}). In follow-up observations by the Whipple collaboration, it became the first AGN found at TeV energies~(\cite{mrk421-whipple}), and in fact the {\em only} of the EGRET detected at TeV energies at the time, providing the first indications of the attenuation of TeV $\gamma$ rays by pair absorption with the extragalactic background light. Mrk~421 was detected by MILAGRO, with a $7.1\sigma$ significance on 2.5~years of data taken between September 2005 and March 2008~(\cite{mrk421-milagro}).  The HAWC light curve of Mrk421, including 387 selected transits, represents the first daily monitoring of a TeV blazar with the sensitivity to actually detect flaring on a single transit. The data are inconsistent with a constant flux with $p-{\rm value}\la 10^{-10}$. Frequent high states of fluxes larger than 1 or 2~Crab have been observed. Recently, HAWC, FACT and {\em Swift}-XRT reported the joint observation of a four day activity of Mrk~421, which reached 2~Crab units above 1~TeV~(\cite{mrk421-flare-hawc}).

\begin{figure} \begin{center}
\includegraphics[width=\hsize]{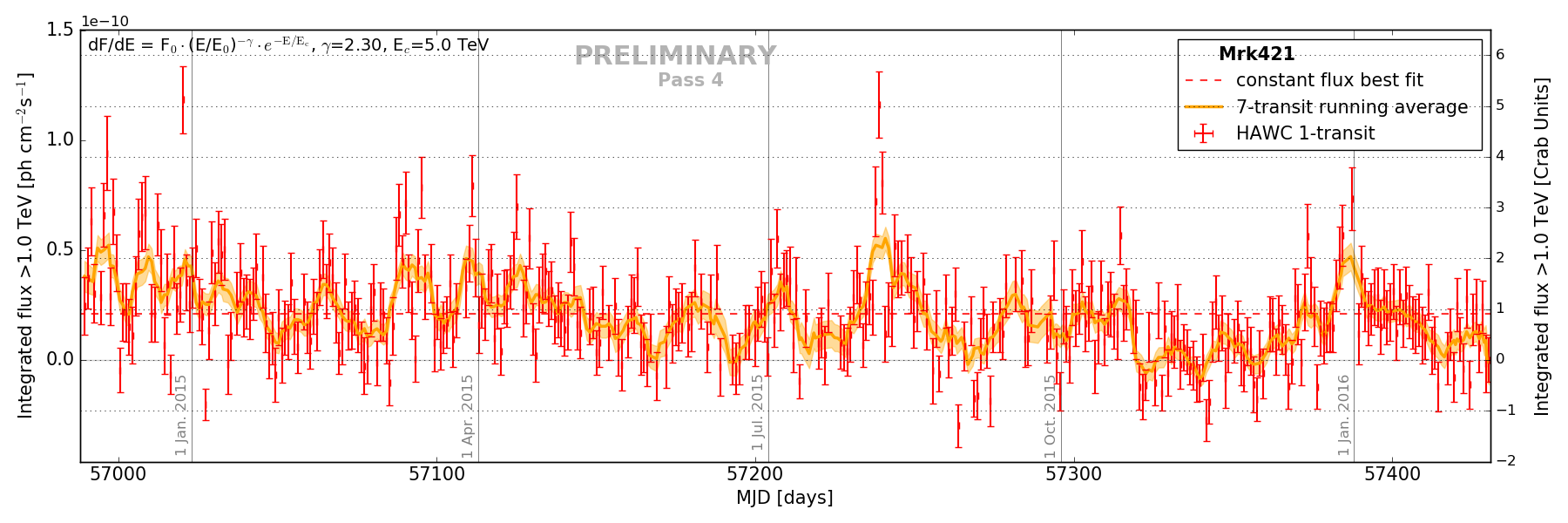}
\caption{HAWC light curve of the BL Lac object Mrk 421 spanning from November 2014 to February 2016. Five periods of three months are indicated by the vertical grey lines. High states, in which the TeV flux of Mrk~421 was stronger than that of the Crab, are clearly noticeable.}\label{lc421}
\end{center} \vspace*{-2mm}
\end{figure}

\begin{figure} \begin{center}
\includegraphics[width=\hsize]{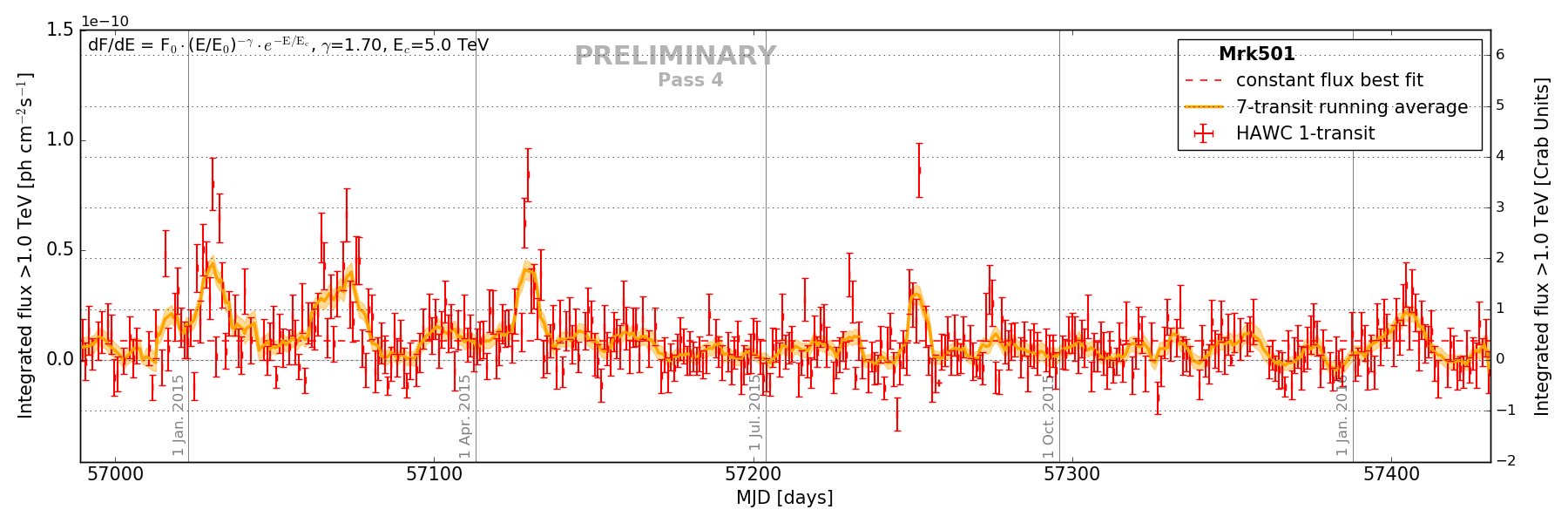}
\caption{Daily HAWC light curve of Mrk 501 between November 26, 2014 and February 12, 2016. The fluxes are calculated assuming a spectral index of 1.7 and an exponential cut-off at 5~TeV. Episodes of flaring activity, with fluxes up to $3\times$ the Crab are clearly seen.}\label{lc501}
\end{center} \end{figure}

At a distance $d_{L}=143~\rm Mpc$, deduced from its redshift $z=0.033$, Markarian 501 is the second nearest known BL Lac object. With a GeV luminosity lower than Mrk~421, it remained undetected all through the {\em CGRO} mission~(\cite{3eg}), despite that ACTs discovered its TeV emission (\cite{mrk501-whipple}). Its GeV detection was achieved by {\em Fermi}-LAT, that has now measured it up to the 0.585-2.0~TeV band, as reported in the 2FHL catalog~(\cite{2fhl}). Mrk~501 was clearly detected with HAWC in the 17-month dataset, allowing to generate the corresponding daily light curve of 390~transits. The light curve, shown in Figure\ref{lc501}, is inconsistent with a constant flux with $p\la 10^{-10}$. More recently, on the 6$^{\rm th}$ of April 2016, the online monitor system of HAWC measured a two-day flare from this object, which reached a flux 2.2 times that of the~Crab~(\cite{mrk501-flare-hawc}).

HAWC has proven its capability to provide timely alerts to the astronomical community. While these two Markarian galaxies have been observed by HAWC with high significance, even daily detections, the current sensitivity of the observatory below 1~TeV has impeded the detection of more distant and less luminous sources, for which the $\gamma\gamma$ absorption of TeV $\gamma$ rays by the extragalactic background light (EBL) becomes relevant. Mrk~421 and 501 are detected up to at least 10~TeV, providing fine probes to study the EBL; in addition, HAWC keeps searching for the long term averaged emission and for flaring episodes of TeVCat and 2FHL sources with $z<0.3$, in order to set limits or detect their luminosities and flaring activity at TeV energies.

\subsection{The HAWC search for gamma-ray bursts}
Gamma-ray bursts (GRBs) are believed to be caused by stellar collapse, in the case of long-duration events, or by the merger of two compact objects, like two neutron stars, in the case of short GRBs. These processes can produce stellar mass black holes. While GRBs are particularly strong at MeV energies~(\cite{grb-comptel}), their exploration of GeV band had to wait for the advent of {\em Fermi}-LAT, which has reported at least 7~GRBs with $E>30~\rm GeV$ photons in 8 years; the highest energy GRB photon recorded so far stands at 95~GeV, from GRB~130427A~(\cite{grb130427a-fermi}). 

HAWC has the capability to detect TeV emission from GRBs~(\cite{grbsens-hawc}). This can be done through the main data acquisition system or with the scaler systems. HAWC is currently performing both triggered and untriggered searches. Of the seven GRBs reported by {\em Fermi}-LAT with photons above 30~GeV, only occurred after November 2014, when HAWC started taking data with 250~WCDs; however, with a declination $\delta = +76^\circ$, GRB~160509A was out of the reach of HAWC~(\cite{grb160509a-hawc}). Prior to full operations, HAWC provided limits for GRB~130427A~(\cite{grb130427a-hawc}).

\subsection{Galactic black holes}
Black holes in X-ray binary systems are known Galactic VHE $\gamma$-ray emitters~(\cite{tevcat}); consistent with being young systems, most of them are located along the Galactic Plane, where HAWC has detected about 40 sources, of which $\sim 25\%$ were previously undetected. A large fraction of the HAWC sources appear to have extended emission and can be positionally associated with objects like SNR~(\cite{gplane-hawc}). Four TeV binary systems are in the declination range of HAWC: LSI +61~303, HESS~J0632+057, HESS~J1832--093 and LS~5039. 
% Respective Galactic locations: 135.67+1.11;  205.66-1.44; 22.48-0.18; 16.90-1.28.
% 
While LSI+61~303 and HESS~J0632+057 are clearly undetected, both HESS~J1832--093 and LS~5039, are located too close to strong HAWC Galactic Plane sources, and their search is affected by source confusion.

The Galactic Center hosts a $4\times 10^{6}~\rm M_\odot$ black hole, which has been put forward by HESS observations as the preferred source of PeV cosmic rays in our Galaxy, produced through hadronic processes~(\cite{gc-hess}). Unfortunately, the HAWC access to the Galactic Center, which culminates at $46^\circ$ from the zenith at Sierra Negra, limits its sensitivity for this relevant source. As the response of HAWC at large zenith angles is investigated, the data currently collected may makes it feasible to have its eventual detection at multi-TeV energies.

\subsection{Primordial black holes}
Primordial black holes (PBHs) are potential remnants of the very early Universe, created as results of topological defects, domain wall / cosmic string collisions~(\cite{pbh-creation}). According to~\cite{bh-emission}, PBHs should radiate thermally with a temperature $T_{bh}\sim 10^{-7}\,{\rm K}~M^{-1}$, in a timescale $\tau \sim 10^{64}\,{\rm yr}~M^{3}$, with $M$ measured in Solar units. Black holes of mass $\sim 10^{14}\,\rm g$ would evaporate in a timescale equivalent to the current age of the Universe, with $T\sim 10^{12}\,\rm K$. As the Hawking radiation rapidly becomes an important mass-loss mechanism, the black hole is set to evaporate in a catastrophic event. HAWC can set the most stringent limits for PBHs evaporating on a wide range of timescales, from 1~ms to 100s, going down to evaporation rates $\la 10^{4}\, \rm pc^{-3}yr^{-1}$, an order of magnitude below the current MILAGRO limits~(\cite{pbh-milagro}).

\subsection{Follow-up on gravitational events from black hole mergers}
The Laser Interferometer Gravitational-Wave Observatory (LIGO) has opened a fresh new field of astrophysical research with the spectacular discovery of the black-hole merger event of the 14$^{\rm th}$ of September 2015, GW150914, very likely located in the Southern hemisphere~(\cite{gw1-ligo}). On December 2016, LIGO detected another clear gravitational-wave signal, whose localization contours intersected the field of view of HAWC at the time of the merger observation~(\cite{gw2-ligo}). Even though the localization uncertainty of the GW151226 event is very large, the precise time information allowed for a search of the event on a very specific time window $\pm 10\,\rm s$ within the HAWC data, with null results~(\cite{gw-hawc}).

Since 2014 the HAWC has worked in coordination with the LIGO/VIRGO observatories to search for high energy electromagnetic counterparts of gravitational wave events. The large field of view and high duty cycle make HAWC suitable for multi-frequency and messenger follow-ups. The next LIGO run (O2) is due to start at the time of this writing and last about six months. The run will feature not only improvements in the performance of LIGO, but also the eagerly expected addition of VIRGO that will allow for localization errors of $\la 20~\rm deg^2$, fitting well inside the HAWC field-of-view~(\cite{ligo-plan}).

\section{Summary and next steps}
The HAWC $\gamma$-ray observatory is in full operation, surveying on daily basis the most energetic phenomena in the sky. Its observations are relevant for our knowledge of black holes on different mass regimes, ranging from the supermassive black holes powering Mrk~421 and Mrk~501, to the $10^{14}\,\rm g$ primordial black holes that may have formed in the very early Universe. As the data accumulate, HAWC will be providing a deeper and deeper view of the sky, probing the extreme physical processes powered by astrophysical black holes.

\section*{Acknowledgements}
The High Altitude Water Cherenkov $\gamma$-ray observatory is a collaboration of over thirty institutions in Mexico and the United States, with recent additional participation of Germany, Poland and Costa Rica. HAWC has been possible thanks to the generous support of the National Science Foundation, the US Department of Energy and the Consejo Nacional de Ciencia y Tecnolog\'{\i}a in Mexico.
% usual and Red HAWC Proyecto CONACYT 271737.

This research has made use of the TeVCat online source catalog (tevcat.uchicago.edu); of the SIMBAD database, operated at CDS, Strasburg, France~(\cite{simbad}); and of NASA's Astrophysics Data System~(www.adsabs.harvard.edu).


\begin{thebibliography}{}
%
\bibitem[Aab \etal\ 2014]{auger-limit} {\em The Pierre Auger and Telescope Array Collaborations:} Aab, A. \etal\, 2014, \textit{ApJ} 794, 172. % Auger photon limit
%
\bibitem[Abbott \etal\ 2013]{ligo-plan} {\em LIGO Scientific Collaboration and Virgo Collaboration}, Abbott, B.P., \etal\ 2013, \textit{ArXiV} 1304.0670v3. % GW plans
%
\bibitem[Abbott \etal\ 2016a] {gw1-ligo}  {\em LIGO Scientific Collaboration and Virgo Collaboration}, Abbot, B.P., \etal\ 2016, \textit{Phys. Rev. Lett} 116, 061102. % GW150916 by LIGO
%
\bibitem[Abbott \etal\ 2016b]{gw2-ligo} {\em LIGO Scientific Collaboration and Virgo Collaboration}, Abbot, B.P., \etal\ 2016, \textit{Phys. Rev. Lett} 116, 241103. % GW151226 by LIGO
%
\bibitem[Abdo \etal\ 2014]{mrk421-milagro} {Abdo, A.A., \etal\ } 2014, \textit{ApJ} 782, 110. % Mrk421 by MILAGRO
%
\bibitem[Abdo \etal\ 2015]{pbh-milagro} {Abdo, A.A., \etal} 2015, \textit{Astrop. Phys.} 64, 4.% PBH limit by MILAGRO.
%
\bibitem[Abeysekara \etal\ 2012]{grbsens-hawc} Abeysekara, A.U., \etal\ 2012, \textit{Astrop. Phys.} 35, 641. % GRB sensitiviy HAWC
%
\bibitem[Abeysekara \etal\ 2015]{grb130427a-hawc} Abeysekara, A.U., \etal\  2015, \textit{ApJ} 800, 78. % GRB 130427A HAWC paper
%
\bibitem[Acero \etal\ 2015]{3fgl} Acero, F., \etal\ 2015, \textit{ApJS} 218, 23. % 3FGL
%
\bibitem[Ackermann \etal\ 2014]{grb130427a-fermi} Ackermann, M., \etal\ 2014, \textit{Science} 343, 42. % GRB 130427A by Fermi
%
\bibitem[Ackermann \etal\ 2016]{2fhl} {Ackerman, M., \etal\ } 2016, \textit{ApJS} 222, 5. % 2FHL catalog
%
\bibitem[Aharonian \etal\ 2006]{crab-hess} Aharonian, F., \etal\ 2006, \textit{A\&A} 457, 899. % HESS Crab Nebula
%
\bibitem[Aharonian \etal\ 2013]{gps-hess} Aharonian, F., \etal\ 2013, \textit{ArXiV} 1307.4690 from ICRC 2013. % HESS Galactic Plane survey
\bibitem[Aharonian \etal\ 2016]{gc-hess} Aharonian, F., \etal\ 2016, \textit{Nature} 531, 476. % HESS Galactic Center Pevatron.
%
% \bibitem[Atkins \etal\ 2001]{milagro} Atkins, R., \etal\ 2001, preprint - astro-oh/0110513. % The MILAGRO instrument.
%
\bibitem[Atkins \etal\ 2003]{crab-milagro} Atkins, R., \etal\ 2003, \textit{ApJ} 595, 803. % MILAGRO detection of M1.
%
\bibitem[Biland \etal\ 2016]{mrk421-flare-hawc} Biland, A., \etal\ 2016, \textit{The Astronomer's Telegram} 9137. 
%
\bibitem[Carr 2005]{pbh-creation} Carr, B.J., 2005, \textit{Proc. of ``Inflating horizon of particle astrophysics and cosmology"}, astro-ph/0511743. % PBH origin
%
\bibitem[Fichtel \etal\ 1994]{grocat} {Fichtel, C.E., \etal\ } 1994, \textit{ApJS} 94, 551.% First CGRO Catalog
%
\bibitem[Galbraith \& Jelley 1953]{cr-atmospheric} Galbraith, W., Jelley, J.V 1953, \textit{Nature} 171, 349. % Light pulses from CRs
%
\bibitem[Genzel \etal\ 1996]{bh-gc} Genzel, R., \etal\ 1996, \textit{ApJ} 472, 153. % The BH in the center of the Galaxy.
%
\bibitem[Hanlon \etal\ 1994]{grb-comptel} Hanlon, L., \etal\ 1994, \textit{A\&A} 285, 16. % GRBs with Comptel
%
\bibitem[Hartman \etal\ 1999]{3eg} Hartman, R.C., \etal\ 1999, \textit{ApJS} 123, 79. % The 3EG catalog
%
\bibitem[Hawking~(1974)]{bh-emission} {Hawking, S.W.} 1974, \textit{Nature} 248, 30. % BH Hawking emission
%
\bibitem[Hui 2016]{gplane-hawc} Hui, M. 2016, \textit{APS April Meeting}, U4.001. % HAWC Galactic Plane.
%
\bibitem[Lennarz 2016]{grb160509a-hawc} Lennarz, D., for the HAWC Collaboration, 2016, \textit{GCN Circular} 19423. % GRB 160509A by HAWC.
%
\bibitem[Oppenheimer \& Snyder 1939]{blackholes2} Oppenheimer, J.R., Snyder, H. 1939, \textit{Phys. Rev.} 56, 455.
%
\bibitem[Ozernoi \& Chertoprud 1966]{agn-variability} Ozernoi, L.M., Chertoprud, V.E. 1966, \textit{Sov. Astron.} 10, 15.  % Early evidence for fast variability in AGN.
%
\bibitem[Punch \etal\ 1992]{mrk421-whipple} {Punch, M. \etal} 1992, \textit{Nature} 358, 477. % Mrk421 Whipple 
%
\bibitem[Quinn \etal\ 1996]{mrk501-whipple} {Quinn, J. \etal} 1996, \textit{ApJL} 456, 83% Mrk501 Whipple
%
\bibitem[Salpeter 1964]{bh-agn} Salpeter, E. 1964, \textit{ApJ} 140, 796. % SMBH in AGN
%
\bibitem[Sandoval, Lauer, Wood 2016]{mrk501-flare-hawc} Sandoval, A., Lauer, R., Wood., J, for the HAWC Collaboration, 2016, \textit{The Astronomer's Telegram} 8922. 
%
\bibitem[Sch\"onfelder \etal\ 2000]{comptel-cat} Sch\"onfelder, V., \etal\ 2000, A\&A % COMPTEL catalog
%
\bibitem[Schwarzschild~1916]{blackholes1} Schwarzschild, K. 1916, \textit{Sitzungberichte der K. Preussischen Akademie der Wissenschaften zu Berlin} 1, 189. % black holes
%
\bibitem[Wakely \& Horan 2008]{tevcat} Wakely, S.P., \& Horan, D., 2008, ICRC 3, 1341. % The TeVCat reference.
%
\bibitem[Webster \& Murdin 1972]{cyg-x1} Webster, B.L., Murdin, P. 1972, \textit{Nature} 235, 37. % discovery of Cygnus X-1
%
\bibitem[Wenger \etal\  2000]{simbad} Wenger, M., \etal\  2000, \textit{A\& AS} 143, 9. % SIMBAD
%
\bibitem[Wilson \etal\ 1963]{haverah-park} Wilson, J.G., \etal\ 1963, \textit{ICRC} 4, 27. % Haverah Park
%
\bibitem[Wood 2016]{gw-hawc} Wood, J. 2016, \textit{GCN Circular} 19156. % HAWC on GW 151226
%
\end{thebibliography}
\end{document}